# In-situ study of creep in Sn-3Ag-0.5Cu solder

Tianhong Gu[1]*, Vivian Tong[1,2], Christopher M. Gourlay[1], and T. Ben Britton[1]

1. Department of Materials, Imperial College London, SW7 2AZ. UK
2. Now at: National Physical Laboratory, Hampton Rd, Teddington TW11 0LW, UK

*Corresponding author: t.gu15@imperial.ac.uk, +44 20 7594 2634

## Abstract

The creep behaviour and microstructural evolution of a Sn-3Ag-0.5Cu wt.% sample with a columnar microstructure have been investigated through in-situ creep testing under constant stress of 30 MPa at ~298 K. This is important, as 298 K is high temperature within the solder system and in-situ observations of microstructure evolutions confirm the mechanisms involved in deformation and ultimately failure of the material. The sample has been observed in-situ using repeat and automatic forescatter diode and auto electron backscatter diffraction imaging. During deformation, polygonisation and recrystallisation are observed heterogeneously with increasing strain, and these correlate with local lattice rotations near matrix-intermetallic compound interfaces. Recrystallised grains have either twin or special boundary relationships to their parent grains. The combination of these two imaging methods reveal one grain (loading direction, LD, 10.4 ° from [100]) deforms less than the neighbour grain 2 (LD 18.8° from [110]), with slip traces in the strain localised regions. In grain 1, $(1\bar{1}0)[001]$ slip system are observed and in grain 2 $(1\bar{1}0)[\bar{1}\bar{1}1]/2$ and $(110)[\bar{1}11]/2$ slip systems are observed. Lattice orientation gradients build up with increasing plastic strain and near fracture recrystallisation is observed concurrent with fracture.

**Keywords:** Pb-free solder, Creep, In-situ EBSD, Recrystallisation and Microstructure evolution.

# 1. **Introduction**

For economical and sustainable success of electronic components, estimating the lifetime of solder components is critical. Solders are used to electronically connect components and their structural integrity is important in maintaining this connection. Due to the local geometry of the joint, solders can be subjected to creep deformation. Creep deformation is a time-dependent plastic deformation mode where the stress is below the yield point. Excessive creep can lead to component distortion and can lead to fracture of the solder joint, reducing service lifetime. It is known that Pb-free solders are hot working metals ($\frac{T}{T_M} \geq 0.6$ [1]), and therefore creep performance is an important material selection parameter for these high temperature applications. Sn-3Ag-0.5Cu wt.% (SAC305) has become the most widely used commercial Pb-free solder alloys in electronics such as servers and computers, and yet the mechanisms of creep failure in these alloys remains uncertain.

In SAC305, three stage creep has been reported [2-6] to which the slope at the steady stage (stage II) of the curve is referred to as the creep strain rate. In primary creep, the initial deformation takes place rapidly and the creep strain rate decreases because of strain hardening, which occurs due to dislocation multiplication during plastic deformation. Creep reaches the secondary stage with a nearly constant creep strain rate when the strain hardening and dynamic recovery compete and the stress exponent obtained from plotting secondary creep rate with stress indicates that dislocations can move through or around obstacles such as grain boundaries and intermetallic compounds (IMCs) by cross slip, climb and thermally activated slip [7]. Finally, in tertiary stage creep, the creep strain rate increases with significant accumulated damage. An effective reduction in cross-sectional area is often found due to necking and/or internal void formation [4, 8-10].

Plastic deformation can occur via different slip systems in the β-Sn matrix [11-22], and the activation of β-Sn slip systems is typically related to the crystal orientation, local stress state, and the ratio of the critical resolved shear stress ($\tau_{CRSS}$), which modifies the (effective) weighted Schmid factor (m) when determining the likely active slip systems. However, a large variation has been obtained due to the influence of temperature history, composition, and orientation of the solder samples [13, 16, 23].

There are many families of slip systems that have been identified in Sn, but the relative activity of these different families of slip systems is only partially understood. Zhou et al. [12], Bieler and Telang [23], and Zuo et al. [17] showed that slip in the [001] and [111] directions on and {110} type planes are the most active and are frequently observed in thermomechanical deformation, while activation in the slip of [100] and [101] directions was relatively hard [12]. Bieler and Telang [23] carried out shear lap tests for 0.8 shear deformation at strain rate of 0.1 s$^{-1}$ using an eutectic Sn-3.5Ag/Cu solder to join two pieces of Cu half-dog-bone at 25 °C. They concluded that the (010)[101] slip system was activated and the slip systems in {101}<10$\bar{1}$> and {121}<10$\bar{1}$> families were very difficult to be activated for a sample with its [001] direction near the loading direction. Zuo et al. [17] performed thermomechanical fatigue tests for SAC307 solder joints and discovered that the (110)[1$\bar{1}$1]/2 is the activated slip system. Slip traces were observed in the region with large $Cu_6Sn_5$, which resisted the development of low angle grain boundaries (LAGBs). In addition, it has been found that the $\tau_{CRSS}$ of the slip systems in the {110}<1$\bar{1}$1>/2 family decrease exponentially with increasing temperature [22] which means that high temperature was required to overcome a certain energy barrier so as to initiate these slip systems. However, Matin et al. [15, 16] reported that the slip systems ($\bar{1}$10)[111]/2, (1$\bar{1}$0)[001], (121)[$\bar{1}$01] and (121)[$\bar{1}$01] were activated for Sn-3.8Ag-0.7Cu bulk samples under mechanical [16] and thermomechanical [15] fatigue deformation. Slip traces and microcracks were observed predominantly in the eutectic regions. It is noted that the applied $\tau_{CRSS}$ values for Matin et al. [15, 16] are different to Zamiri et al. [12, 17, 22, 23], who consider different crystal orientations for single-crystal samples under uniaxial tensile test from ref [12]. Matin et al. [15] selected 'consistent' values of $\tau_{CRSS}$ for families, assuming a constant strain rate at 293 K and for family 1, 2, 9 and 10, these were extracted from their experimental measurements from single-crystal experiments of pure β-Sn (99.99%) performed in ref [16] and family 3 - 5 are estimated values by correlating the slip systems with the atomic line density (Peierls model based). The CRSS ratios are given in Supplementary Table S1.

Furthermore, the dependency of orientation of the solder joint on activation of slip systems is reported in the literature [13, 20, 23, 24]. Zhou et al. [13] stated that thermally cycled SAC305 solder joints with c-axis nearly parallel to the chip-side interface showed more slip activity than samples with c-axis perpendicular to the chip-side interface. They also reported that the recrystallisation generating new grains was caused by gradual lattice rotation, which was observed around the Sn <110> axes and formed as a result of the easier activation of slip systems in the {110}<001> family.

By carrying out a first principle atomistic study and uniaxial tensile tests, Kinoshita et al. [20] and Dong et al. [19] found that samples with an orientation close to [001] had slip in the (101) plane activated easily for pure β-Sn and SAC305. For the [100] and [110] orientations, the (110) and (100) + (010) planes are easier to slip [20].

However, there is not only a very limited number of works supporting the used $\tau_{CRSS}$ values, but also a lack of a systematic study on the $\tau_{CRSS}$ values for β-Sn. This is because most works consider the microstructure within these alloys as a homogenous phase [12, 15-17, 22, 23, 25-27] and yet we know it contains regions of primary β-Sn and eutectic regions consisting of β-Sn with embedded IMCs.

Some works have studied microstructural evolution of the solder during in-situ creep deformation [2, 3, 28-32]. The main observations are an increase in surface roughness [2, 28], the formation of shear bands [2, 30], growth of subgrains [2, 3, 29-32] and dynamic recrystallisation [2, 30] with continuous straining. Tian et al. [30] also reported that the creep deformation of each sample varies due to the different grain orientations (due to the anisotropic properties of β-Sn). Among them only Zhang et al. [3, 31, 32] has used in-situ EBSD to improve the understanding of dynamic deformation of Sn-Cu and Sn-Ag solder joints during creep tests. They found that at low strain rate of $1\times10^{-4}\,s^{-1}$, a significant strain concentration was observed for Sn-Ag/Cu solder joints at the Sn-$Cu_6Sn_5$ interface with the formation of shear bands, while a relatively uniform deformation was introduced with wave-like deformation bands formed for Sn-4Ag/Cu solder joints. The grains deformed with lattice rotation, which depended on their orientation, leading to grain subdivision and grain boundary migration, however, for these tests recrystallisation was not observed [32]. Furthermore, the samples used in their work [3, 31, 32] have different microstructures, namely orientations and number of grains, which will have a significant effect on the creep behaviour.

In previous work the twin nucleated recrystallisation has been suggested [33-36], however there has been no direct evidence of this provided, as without in-situ observations the mechanisms are unclear. In the work of Arfaei et al. [33], the correlation of solder fatigue with both recrystallisation and intergranular crack growth was observed for SAC305 samples that had undergone in thermal cycling. They observed that recrystallisation started to form in the highly-strained regions such as in the bulk solder near the Si chip-side and close to coarsened precipitates, likely from the spread in orientation around LAGBs and twin boundaries. In the work of Telang et al. [34], they reported continuous recrystallisation in a Sn-3.5Ag joint during thermomechanical fatigue. Based on ex-situ EBSD observations, they found that some of the newly recrystallised orientations were twin orientations to the initial orientations and the {110} plane slip traces were generated simultaneously with increasing recrystallisation. Furthermore, Mattila et al. [35] tested Sn-Ag-Cu solder joints under thermomechanical loading and demonstrated that the twin nucleated recrystallisation can appear preferentially at grain or phase boundaries and near solder / substrate interface to achieve less energy consumption.

To address these complexities, here we perform an in-situ study of creep in controlled SAC305 microstructures, with a focus on when and where grain polygonisation and recrystallisation occur, with respect to grain boundaries and eutectic regions where the β-Sn contains embedded $Ag_3Sn$ and $Cu_6Sn_5$ IMCs, which can deform differently during creep [37]. An in-situ study also removes ambiguities over features that could develop as the material relaxes after unloading due to room temperature creep and recovery.

In this work, electron microscopy is combined with in-situ deformation. This enables us to follow the evolution of slip and recrystallisation, via changes in lattice orientation. Fast imaging with electron channelling contrast imaging (ECCI) enabled us to confirm that recrystallisation occurred during the deformation process, and EBSD enabled us to understand the relationships of new grains with respect to the parent microstructure.

## 2. Experimental procedure

The creep tests were performed on SAC305 large scale dog-bone samples that had been solidified directionally to produce columnar grains along the gauge length with controlled orientation and lengthscale similar to ref [37]. The dog-bone shape was first cut using electrical discharge machining (EDM) from a flat cold-rolled SAC305 sheet of 1.5 mm thickness into 10 (gauge length) × 2 × 1.5 mm dimensions (Figure 1a). This was then melted and solidified using Bridgman growth, with a pulling rate of 20 μm s$^{-1}$ in a temperature gradient of 3 K mm$^{-}$ [1], to fabricate the sample with a controlled microstructure (length scale and orientation) as shown in Figure 1b.

Since EBSD is a surface sensitive technique, a high-quality surface finish was prepared by mechanical polishing with 0.05 µm colloidal silica and then improved using broad ion beam milling at room temperature with 3 keV and 3° and followed by 3 keV and 1° for 3 hours (Gatan PECS II) [37].

The creep tests were performed with a 2 kN Gatan Mtest 2000E tensile stage (Figure 1b) at room temperature (~298 K) and constant stress of 30 ± 2 MPa inside a Quanta FEI 450 SEM. Two tests were carried out in tensile mode under load control until sample fracture (bi-crystal sample) without interruption and to 3% strain (single crystal sample) with interruption at every 0.5% strain interval. During microstructural mapping, load was held without unloading or reducing and the absolute extension of the samples was recorded to measure the strain (the strain measured is the engineering strain which is determined by extension divided by length). The microstructural deformation of the sample was observed in-situ using ECCI with repeating forescatter diode (FSD) imaging and EBSD-mapping with a Bruker e-Flash$^{HD}$ EBSD detector.

The bi-crystal sample was observed in a relatively large region of interest (ROI) which contains grain boundaries and dendrite and eutectic regions. To ensure that deformation was localised to the field of view, the sample was mechanically back thinned by diamond pen for several micrometres in the centre of the gauge as shown in Figure 1(a). The single crystal sample was focused within the eutectic region using a smaller ROI and the microstructural deformation was observed by EBSD-mapping until 3% strain with a 0.5% strain interval.

Scripting in Bruker ESPRIT 2.2 was used to create a routine to enable repeat FSD and EBSD mapping, so called "auto-FSD and auto-EBSD". The FSD images were captured with the detector retracted (to reduce topography contrast and increase electron channelling contrast for subtle orientation changes) with a detector distance of 40 mm and a 1.2 µm step size (1200 × 799 pixels), where each image took 1 minute to capture. After the FSD imaging, the detector was moved automatically to the EBSD position (~10 mm) and the EBSD scan started as soon as the detector reached the set position. The EBSD maps were captured at a 5 × greater step size than the FSD image, leading to a step size of 6 µm for the bi-crystal sample (larger ROI). This significantly reduces the mapping time to measure the orientation information of the sample. The duration of each auto-FSD and auto-EBSD pair was approximately 7 minutes. Likewise, the EBSD maps for the single crystal sample (smaller ROI) were captured at 0.2 µm step size and each auto-FSD and auto-EBSD pair took approximately 60 minutes. The initial microstructure and orientation of the gauge for the sample were scanned using a large-scale EBSD map before with step size of 12 µm inside the load frame before loading. After that, the creep test was performed together with auto-imaging. The microstructure of the sample was compared before and after deformation. The EBSD data was analysed with post-processed using the "absorb surrounded zero solutions" function in Bruker ESPRIT 2.2.

## 3. Results and Discussion

### 3.1. Creep strain curves

Figure 2 shows the creep strain curves of a bi-crystal and a single crystal sample tested at constant stress (~30 MPa) and temperature (~298 K) until a creep strain of 32% (until fracture) (Figure 2a) and 3% (until early secondary stage creep) (Figure 2c) respectively. Overall, the sample deforms with three stages corresponding to different stages of creep deformation (Figure 2a). The change in creep strain rate is illustrated by the creep strain rate (% $s^{-1}$) vs. strain curves (Figure 2b). In the primary stage creep, the strain increases rapidly due to dislocation glide, multiplication, and formation of substructure. When the deformation enters the secondary stage, the creep strain rate decreases to a nearly constant value and it occupies most of the creep lifetime. With further deformation, the creep strain rate starts to increase again leading to the failure of the sample (Figure 2a). The strain at the onset of secondary and tertiary stage creep can be determined from Figure 2b, with values of 1.9 and ~10% respectively (a linear relationship with gradient close to 0 is achieved). The secondary creep strain rate is on the order of $10^{-4}$ % $s^{-1}$, which is consistent with the finding in ref [37]. Figure 2c shows an inverted primary creep and the strain rate increases with strain. This is consistent with solute drag mechanisms [7], where the initial dislocation density is low and viscous in the sample, and the dislocations interact with each other to develop density, implying that the primary stage is controlled by dislocation climb and / or glide. Once a stable substructure is reached (~ 2% strain here), the density of mobile dislocations is lowered by creep straining and results in a gradual decrease in strain rate [38, 39]. The corresponding microstructural evolution at each stage of creep is explored in-situ in the next section.

### 3.2. Microstructural evolution of the bi-crystal sample

The sample forms a microstructure with β-Sn dendrite and intermetallic-containing eutectic regions (Figure 3). The measured β-Sn dendrite arm spacing is ~50 ± 3 μm. The interdendritic eutectic contains IMC particles with size and spacing of ~0.5 ± 0.1 and 1.18 ± 0.02 μm considering $Ag_3Sn$ and $Cu_6Sn_5$ particles together measured from micrographs such as in Figure 3. Note that a similar phenomenon can sometimes occur in SAC305 ball grid array (BGA) solder joints [40].

The sample has two β-Sn domains (grains) with the loading direction aligned 10.4° from a <100> in the upper grain 1 eutectic regions and 18.8° from a <110> in both the upper and lower grain 2 dendritic regions (Figure 4). The dendritic β-Sn in grain 2 has a blue colour in the IPF-LD map (Figure 4a), whereas the eutectic β-Sn component in grain 1 and 2 have green and blue colours respectively. The initial microstructure of the sample is schematised in Figure 4c. In the red boxed region, the green eutectic β-Sn has a different orientation than the surrounding blue dendritic β-Sn, while in the lower blue boxed region, the dendritic and eutectic β-Sn have the same orientation.

Figure 4 illustrates the macroscopic deformation of the sample by showing the sample before and after 32% strain. New grains with different crystallographic orientations, shown in yellow (grain 3) in Figure 4d, formed in the regions with high local misorientation (Figure 4e). These regions are mainly located in the β-Sn dendrite (grain 2 (blue)) both at the dendrite-eutectic boundaries and near the grain boundaries due to the heterogeneity in rotation between the two orientations. The dendrite and eutectic regions can be distinguished clearly in Figure 4c. The overall misorientation increases with deformation from the undeformed condition (Figure 4b) to the fracture of the sample (Figure 4e). The pole figures (PF) in Figure 4f show lattice rotation of the two grains before and after deformation, indicating that the blue orientation (indicated with blue circles initially) experiences a greater range of rotations and spreads in orientation more than the green orientation (indicated with green squares initially). The deformed orientation of the green eutectic grain1 (indicated with green dashed squares) rotates slightly in one direction. The initial blue orientation develops into 2 orientations, i.e. the dendritic β-Sn (indicated with yellow dashed circles) and the eutectic β-Sn (indicated with magenta dashed circles) rotate in different directions and result in a high misorientation at the dendrite-eutectic boundary showing as yellow hot spots in Figure 4e. Figure 4 gives the general overview of the evolution of microstructural morphology and orientation of the sample, which is then revealed in detail with Figure 5 – Figure 7 by in-situ FSD and EBSD at different stages of creep in the next sections.

### 3.2.1. Primary and early secondary stage creep

The microstructural evolution of the sample is revealed between primary to early secondary creep in Figure 5. The creep curve of the sample is given in Figure 5a. Figure 5b – 5g shows the microstructure of the sample at early secondary creep.

The FSD image at the initial condition shows a large contrast in Figure 5b not only between grains and subgrains but also between the two microstructural regions, i.e. dendrite and eutectic regions, which is a result of the difference in topography (Figure 5b). The two initial orientations, i.e. grain 1 (green) and grain 2 (blue) (Figure 5c) have some regions with slightly higher misorientation, which is associated with the presence of subgrains (indicated with black and white arrows in Figure 5b, 5d, 5e). The high angle grain boundaries (HAGBs) and subgrain boundaries (LAGBs) are illustrated in Figure S2 at each strain level. (A common misorientation angle with respect to 'grain' as identified with EBSD thresholding of grain boundaries at 5° and subgrain boundaries at 2°).

With increasing strain (Figure 5e – 5g), small regions of recrystallised orientation grain 5 (magenta) appears at the boundary between β-Sn dendrite and eutectic regions, where the β-Sn dendrite and eutectic β-Sn have different orientation (indicated with red arrows in Figure 5f). Slight colour change (light yellow) is observed in the β-Sn dendrite at the boundary between dendrite and eutectic region (Figure 5f). These changes in the IPF-LD map are caused by lattice rotation, leading to an increase in misorientation (indicated with yellow and white arrows in Figure 5f – 5g). The main orientations and the morphology of the microstructure in the scanned region show no change between the two stages of creep in Figure 5b & 5e.

### 3.2.2. Secondary stage creep

With continued deformation (Figure 6b – 6j), the grain contrast in the FSD maps reduces significantly due to the increase in surface roughness (Figure 6b, 6e, 6h) which also causes less indexing in the EBSD maps (Figure 6c, 6f, 6i). Shear bands appear gradually in the eutectic regions around IMCs in the FSD maps (indicated with black arrows in Figure 6e, 6h). These shear bands are observed as the slip causes out of plane displacement which increases with increasing total strain as the soft β-Sn phase flows around the hard IMC phase and forms several small bumps.

In addition, new subgrain boundaries are formed (Figure 6c, 6f, 6i and Figure S2) during secondary stage creep due to a gradual increase in misorientation and the high misorientations (indicated with white arrows in Figure 6d, 6g, 6j) are located in the regions with high surface roughness. Due to local lattice rotation, there is a high orientation gradient subjected with the interface as indicated by high misorientation average, especially in the β-Sn along the grain boundary and at the boundaries between dendrite and eutectic regions to form new orientations, such as grain 3 (which becomes yellow), which are annotated with red arrows in (Figure 6c, 6f, 6i). Small recrystallised grains appear increasingly at the grain boundary with magenta (grain 5) and darker blue (grain 6) colours (Figure 6c, 6f, 6i). Table 1 shows the two observed special boundaries for the bi-crystal sample, where the grain boundary misorientation and character (axis/angle) was determined from the EBSD measured orientation data. The magenta recrystallised grain (grain 5) and its parent blue grain share a common <100> direction (Figure 6k) and have a 63° misorientation from each other. The calculated angle deviation is approximately 4.4° to the closest common <100> axis (Table 1). Thus, the recrystallised grain 5 (magenta) likely formed due to deformation twinning [41, 42]. The recrystallised grain 6 (darker blue) appears within the initial green orientation during deformation. Figure 6k shows that the common direction between grain 6 (darker blue) and grain 1 (green) is <001> with a misorientation angle of ~28°, which is a special coincident site lattice (CSL ∑17) boundary according to ref [43]. The angle deviation measured from the closest common <001> axis is 3.0° (Table 1).

| Grain | Angle of special boundary | Measured Misorientation | Common axis | Angle deviation from common axis |
|---|---|---|---|---|
| 2 & 5 | 62° (twin) | 63.1° | [100] | 4.4° |
| 1 & 6 | 28° (CSL) | 28.6° | [001] | 3.0° |

*Table 1. The observed special boundaries with the common axis and angle deviation from the common axis determined for the bi-crystal sample. The twin and coincident site lattice (CSL) boundaries are identified and labelled according to ref [43].*

### 3.2.3. Tertiary stage creep

In tertiary creep, the creep strain rate increased leading to acceleration of the microstructural deformation and fracture of the sample. There is a significant increase in surface roughness and shear bands are observed in the FSD maps (Figure 7).

During the transition from secondary to tertiary creep, shear bands start to progress across the grain boundaries in the strain localised regions to form a microcrack, where high surface roughness is present (Figure 7b). The microcrack then grows into a long crack and propagates through the sample in later tertiary stage creep (Figure 7e).

Within the region with β-Sn dendrites and eutectic β-Sn in different orientations (red boxed region in Figure 4c, the grain 1 (green) and grain 2 (blue) evolve differently during creep. The blue orientation (grain 2) exhibits more colour change corresponding to more obvious lattice rotation than the green orientation (grain 1) (Figure 7c, 7f). The misorientation across the map decreases here due to the formation of new undeformed recrystallised grains that releases the stored energy from previous stages (Figure 7d, 7g).

A region with high surface roughness (the highlighted region in Figure 7e) is characterised by high magnification EBSD. The scanned map is located mainly within the eutectic region in the red boxed region in Figure 4c, which has both $Ag_3Sn$ and $Cu_6Sn_5$ phases outlined with green and blue, respectively (Figure 7h) and has a (green) near-[100] orientation (Figure 7i). A large number of subgrains were formed showing as high misorientation in the grain reference orientation deviation (GROD) map (Figure 7j). Highly misoriented new grains were generated within the highly-roughened strain concentrated region and the HAGBs are highlighted with red lines in Figure 7h. Figure 7i shows that the recrystallised grains in the β-Sn dendrite (indicated with a black arrow) have larger size than in the eutectic region (indicated with red arrows). The deformed microstructure has a wavy surface topography with many slip traces (Figure 7e).

### 3.3. Microstructural evolution of the single crystal sample

The microstructural deformation within the eutectic region of the single crystal sample with a near -[110] orientation (deviation angle ~5.5°) in the loading direction is illustrated in Figure 8 with increasing strain during creep. The FSD image (Figure 8a) illustrates the contrast between the IMCs and β-Sn and the topography of IMCs. The IMC phases, namely $Ag_3Sn$ (green) and $Cu_6Sn_5$ (blue) are identified clearly in the phase map (Figure 8b).

At the initial stage of deformation ($\varepsilon_{LD} = 0\%$), the scanned region of the sample shows a clear single crystal of β-Sn in the IPF-LD map (Figure 8c) and high misorientation spots are present in Figure 8d because the pre-deformation occurs easily due to the coefficient of thermal expansion (CTE) mismatch between the β-Sn and IMCs during cooling after solidification.

Recrystallisation happens quite early, at a strain level $\varepsilon_{LD} = 1\%$, and the recrystallised grain is located near the eutectic IMCs as shown in Figure 8e (highlighted with a white circle). With increasing strain, the recrystallised grain grows and new recrystallised grains (1-3) appear as strain increases from $\varepsilon_{LD} = 2\%$, (Figure 8g – 8h) to $\varepsilon_{LD} = 3\%$ (Figure 8i – 8j). The recrystallisation develops in the region which has high misorientation at the previous strain stage (Figure 8d, 8f, 8h, 8j). The average misorientation increases continuously across the map, while slight decrease in local misorientation is observed near the recrystallised grains (highlighted with white, yellow and green circles accordingly in Figure 8d – 8j). The initial deformation is controlled by the length scale, size and spacing of the eutectic IMCs, which is probably caused by IMC pinning of dislocations that causes the formation of subgrains.

Figure 9a - 9b show the zoom-in IPF-LD maps within the circled regions in Figure 8c. The recrystallised grains (1-3) develop continuously until a strain of $\varepsilon_{LD} = 3\%$ and localise in the β-Sn close to the IMCs. Furthermore, the sample orientation shows obvious rotation in the (110) and (100) PFs from $\varepsilon_{LD} = 0\%$ to $\varepsilon_{LD} = 3\%$ (Figure 9e) with a common (001) plane, indicating the [001] is the rotational axis (the plane has the least amount of rotation among the 3 planes) during uniaxial creep test. Figure 9f illustrates the formed recrystallised grains (1-3) share a common plane with its parent grain (blue circles in the PFs), which is highlighted in yellow square. For recrystallised grain 1 and parent grain (blue), the near-parallel (100) planes are clear. The measured misorientation is 61.8° and the angle deviation is 2.9° from the closest common axis [100] given in Table 2, so the recrystallised grain 1 has a twin orientation relationship with its parent grain. The recrystallised grain 2 and 3 have near-parallel (110) planes to the parent grain (blue) with an angle deviation of 2.5° and 3.1° respectively (Table 2). The measured misorientation angles are 23.8° and 8.5° (Table 2), implying the special boundaries CSL $\sum 27$ and $\sum 30$ of β-Sn are formed [43].

| Grain | Angle of special boundary | Measured Misorientation | Common axis | Angle deviation from common axis |
|---|---|---|---|---|
| Parent & 1 | 62° (twin) | 61.8° | [100] | 2.9° |
| Parent & 2 | 22.3° (CSL) | 23.8° | [110] | 2.5° |
| Parent & 3 | 7.3° (CSL) | 8.5° | [110] | 3.1° |

*Table 2. The observed special boundaries with the common axis and angle deviation from the common axis determined for the single crystal sample. The twin and coincident site lattice (CSL) boundaries are identified and labelled according to ref [43].*

### 3.4. Slip system and dislocation activities

During uniaxial deformation, with single crystals and large grains, we assume that the macroscopic boundary condition controls deformation. This allows the propensity for slip to be assessed using the Schmid's Law, where the relative activity of the slip systems can be determined according to their critical resolved shear stress ($\tau_{CRSS}$) and Schmid factor (m), as given in Equation (1).

$$\tau_{CRSS} = \frac{F}{A} \cos\phi \, \cos\lambda \tag{1}$$

where F is the applied load, A the cross-sectional area of the sample, and $\phi$ and $\lambda$ are the angle between the loading direction and the slip plane, and the loading direction and the slip direction respectively. The Schmid factor is $m = \cos\phi \times \cos\lambda$. Note that within the β-Sn system, the slip system with the highest Schmid factor may not activate first, as $\tau_{CRSS}$ varies between slip systems.

For the 32 β-Sn reported slip systems, their likelihood of activity was ranked according to prior work using expected ratios of the $\tau_{CRSS}$ (CRSS-ratio) [12, 21] and the ranking given in Supplementary table S1. The values of Schmid factor were calculated with loading along the [110] and [100] orientations using Matlab with Equation (1). The weighted slip system with minimum $\tau_{CRSS}$ and maximum m will be activated first [16], where the weighted Schmid factor ($m_W$) of the sample was calculated from

$$m_{W^i} = \frac{\tau_{CRSS^i}}{\tau_{CRSS^{Min}}} \quad (2)$$

The values of weighted Schmid factor ($m_W$) for the grain 1 with a near-[110] orientation (Figure 10) and grain 1 with a near-[100] orientation (Figure 11) were ranked from the easiest to hardest for the bi-crystal sample.

The analysis of activated slip systems for the near-[110] orientation (grain 2) in the bi-crystal sample is shown in Figure 10. The experimental slip trace was measured from the FSD image (Figure 10a). Figure 10b labels the highest $m_W$, on the crystal unit cell for the near-[110] orientation. The slip plane (cyan shaded planes), slip direction (green arrowed lines), calculated and experimental slip traces (magenta and blue dashed lines) are labelled on the unit cell, which is plotted from the measured Euler angles. The most active slip system for this crystal orientation is #4 (1$\bar{1}$0)[001] according to its highest value, $m_W$ = 0.391 and the experimental slip trace (blue dashed line in Figure 10b) shows the best match with the calculated slip plane and direction of the slip system. The measured deviation angle (ɵ) between the calculated and experimental slip traces for grain 2 is ~3°. Although the #1 (100)[001] and #9 (1$\bar{1}$0)[111]/2 slip systems have the second and the third highest values of $m_W$, they are unlikely to be active because the calculated slip plane for the #1 (100)[001] slip system does not match the observed slip trace experimentally and the slip distance for #9 (1$\bar{1}$0)[111]/2 is relatively longer (and hence, requires greater force to move) than the slip system #4 (1$\bar{1}$0)[001] for the same diagonal slip plane (see supplementary Figure S1a).

Figure 11 shows the analysis of slip systems for the near-[100] orientated grain 1 in the bi-crystal sample. Two different slip traces were observed and are labelled using blue and red dashed lines in the FSD image (Figure 11a). By considering $m_W$, the highest two values for the near-[100] orientations are shown in the crystal unit cells diagrams (Figure 11b – 11c). #10 ($1\bar{1}0$)[$\bar{1}\bar{1}1$]/2 and #8 (110)[$\bar{1}11$]/2 are determined to be the most active slip system for grain 1, since the two experimental slip traces (the blue and red dashed slip traces) are in line with the calculated slip planes and directions for the two slip systems optimally. The measured deviation angles are ~1 and 2° between the calculated and experimental slip traces for grain 1. The # 11 (110)[$1\bar{1}0$] and #12 ($1\bar{1}0$)[110] slip systems are less favourable because of their relatively longer slip distances than the slip systems for #8 (110)[$\bar{1}11$]/2 and #10 ($1\bar{1}0$)[$\bar{1}\bar{1}1$]/2 for the same diagonal slip planes (Figure S1b).

This analysis shows that the initial green near-[100] orientation deforms at higher strain than the initial blue near-[110] orientation during creep, which results in more volume of recrystallisation generated. It is found that the crack direction is crystallographic. The formed trace of crack (Figure 10a) looks to be consistent with the slip bands, i.e. ($1\bar{1}0$) slip plane of SS #4), where the crack interface is parallel to the blue experimental slip trace. Also, the slip traces are more prominent in the near-[110] grain, implying that the repeat activation and progression of dislocations is easier for the near-[110] orientation than for the near-[100] orientation.

### 3.5 Recovery and recrystallisation mechanisms of solder

Dynamic recovery is demonstrated in this work, where the highly polygonised subgrains have been formed in regions which correlate previously with regions with high crystal misorientation (Figure 8). These subgrains are formed from continuous annihilation and rearrangement of dislocations via dislocation climb and cross-slip (as tin is so hot that thermally activated rearrangement processes are easily achieved). The remaining dislocations form lower energy subgrain boundaries, misoriented a few degrees from the initial crystal orientation. This process is dynamic recovery, where the low angle subgrain boundaries contain stored energy, and the density and misorientation of these boundaries correlates with the total stored energy within the substructure.

Recrystallisation occurs as the deformation continues, as the stored energy can be further reduced by new grains absorbing the dislocation structures. The interfaces between the recrystalised grains and the previous material are HAGBs (Figure 7). The dislocations are removed by nucleation and growth of new grains that eliminates the driving force for further recrystallisation [35, 44, 45]. New grains with high misorientation angle relative to the adjacent matrix are recrystallised grains [46] and they release the internally stored energy by forming HAGBs and increasing the grain boundary area as the new recrystallised grains consume the dislocated matrix. The nucleation of recrystallisation is controlled by stored energy, often in the form of stored dislocations such that the nucleation rate and driving force for nucleation of recrystallisation are proportional to the dislocation density [35, 47-49]. The nucleation of recrystallisation is commonly initiated at dendrite-eutectic and grain boundaries as well as large IMCs, which are energetically favourable sites for nucleation [35, 47]. In the primary tin dendrites, dislocations can accumulate (especially when two slip systems interact) and form substructure, as evident in local lattice misorientation. In the EBSD maps, these interactions are often observed as lattice rotation gradients, and such rotation occurs in <001>, <$\bar{1}\bar{1}1$>/2 and <$\bar{1}11$>/2 directions, corresponding to slip systems on the {110} plane which activate at low stress, consistent with the outcomes in the literature [44, 50, 51]. Moreover, these interactions are typically observed in regions with defects or second phase particles (IMCs) due to the deformation heterogeneity associated with their relative high slip strength [52, 53].

The failure of the samples is observed as propagation of ductile cracks together with polygonisation and recrystallisation in the bulk. During creep, the plastic deformation is firstly localised to form parallel shear bands with high surface roughness because of material flow at beginning of the secondary creep. These shear bands correspond to slip traces associated with highly favoured slip systems that depend on the initial orientation of the grain [3]. Referring to the EBSD maps in Figure 5 and Figure 6, polygonisation and discontinuous recrystallisation are developed continuously at the region where the mismatch of CTE is large and the induced strain is high in comparison to the surroundings, i.e. in the regions near IMC interfaces, dendrite-eutectic boundaries and grain boundaries.

As the strain increases (tertiary creep in Figure 7), the crack initiates in the strain-localised regions as the consequence of deformation heterogeneity. The cracking of solder propagates progressively with simultaneous evolution of microstructure, namely recrystallisation, which becomes more prominent at higher strain levels. Therefore, there is a strong correlation between recrystallisation and crack propagation, i.e. the high-angled recrystallised grain boundaries assist the crack propagating across the solder matrix. The high-angle and high-energy boundaries produced by recrystallisation can facilitate grain boundary sliding and void formation at boundaries, which provide a favourable energy path for cracking, as supported by ref [13, 24, 34, 35, 48, 54].

## 4. Conclusions

The creep behaviour and microstructural evolution of directionally solidified SAC305 solder with single crystal and bi-crystal structures have been investigated through in-situ creep testing under constant stress of 30 MPa at room temperature (~298 K). The following conclusions can be made.

1. During creep, the microstructural deformation evolves as indicated by the evolution of lattice rotation, which is influenced by the initial microstructural morphology and orientation of the sample.

2. The dendritic β-Sn and eutectic β-Sn deform differently during creep, i.e. the length-scale of recrystallisation is much smaller in eutectic regions than in the dendrite and the deformation is more localised in the eutectic region, while the deformation in the dendritic region is relatively extensive. This is in agreement with our prior work on different samples with mechanisms explained in ref [37].

3. The grain with a near-[110] orientation in the loading direction rotated faster and exhibited more prominent slip traces than the grain with a near-[100] orientation in the loading direction. This indicates that propagation of slip in the near-[110] oriented grain is more favourable.

4. In-situ measurements confirm that recrystallisation develops under load, rather than after the test while the sample is left at room temperature. This in-situ work shows that the interpretations made from previous ex-situ works [33-36] are indeed valid, even though there is possibility of significant microstructural change at room temperature.

5. Localised recrystallisation occurs at the early stage of creep within the β-Sn at the grain boundaries, dendrite-eutectic boundaries and close to IMCs because of large heterogeneity in rotation between the orientations, microstructural regions and phases. The formed recrystallised grains have either twin or special boundaries to their parent grains because twin and CSL boundaries have relatively low energy configuration when misorientation increases locally. This observation is consistent with ref [43].

6. Polygonisation starts to develop increasingly at early stage of secondary creep in the strain localised regions such as the regions close to the IMCs, near the dendrite-eutectic boundaries and grain boundaries due to the continuous dislocation formation and rearrangement locally resulting from high heterogeneity at grain and phase boundaries. Since recovery takes place deformation accumulates as indicated with gradual increase in surface roughness and the formation of shear bands during secondary stage creep [35, 53].

7. The microstructural deformation accelerates in the region with high roughness to generate significant subgrain and recrystallisation, leading to crack formation and fracture of the sample at tertiary creep. This is because that the soft β-Sn phase flowed around the hard IMC phase during creep and caused the formation of a wavy morphology with slip bands resulting from crystal slip generated by gradual lattice rotation [50, 51, 53].

8. Activated slip systems were identified using the observed slip traces to measure the slip plane and calculating the geometrically best aligned slip direction with maximum Schmid factor. The $(1\bar{1}0)[001]$ slip system was observed to be activated for grain 2, which has its [110] 18.8° to the loading direction. On the other hand, two slip systems $(1\bar{1}0)[\bar{1}\bar{1}1]/2$ and $(110)[\bar{1}11]/2$ were activated for grain 1, which has its [100] 10.4° to the loading direction. With increasing deformation, more slip traces were formed and became more prominent by increasing the local surface roughness, which initiated cracks that led to fracture.

### Author Contributions:

TG drafted the initial manuscript and conducted the experimental work. VT developed the scripting for the in-situ experiments. CG and TBB supervised the work equally. All authors contributed to the final manuscript.

### Acknowledgements:

TBB would like to thank the Royal Academy of Engineering for his research fellowship. CG would like to thank EPSRC (EP/M002241/11) for funding of his research fellowship. We would like to thank EPSRC (EP/R018863/1) for funding. We thank Dr Sergey Belyakov and Dr Ning Hou for support in the initial fabrication of the samples. We also very thankful to Prof Fionn Dunne, Dr Finn Giuliani, Dr Yilun Xu and Dr Te-Cheng Su for support and discussion in developing the Matlab code of Sn slip systems, and Dr Yi Guo and Ms Ruth Birch for suggestions and discussions in developing Matlab code of Sn crystal orientations. The microscope and loading frame used to conduct these experiments were supported through funding from Shell Global Solutions and provided as part of the Harvey Flower EM suite at Imperial. We acknowledge the reviewer for some aspects of the discussion.

**Figure Captions**

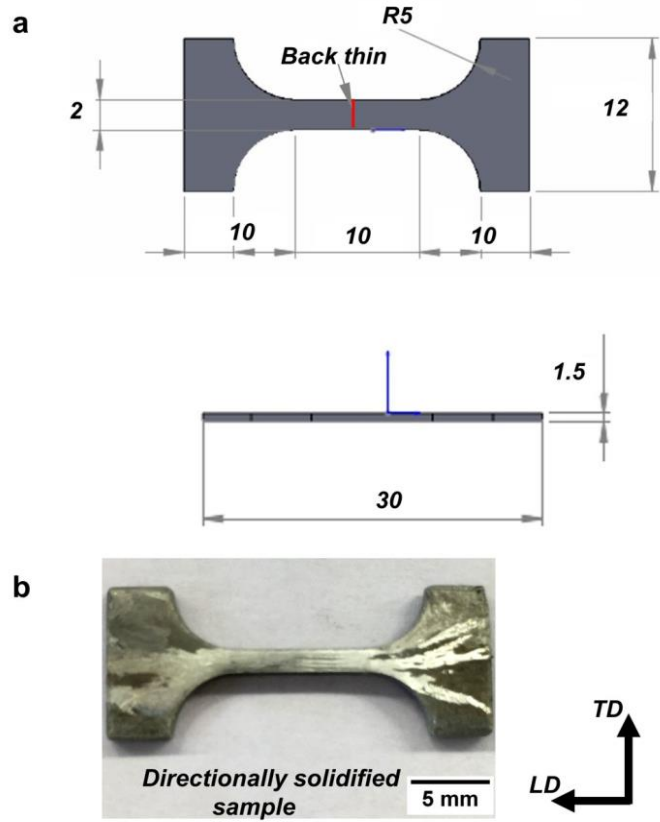

*Figure 1. (a) Schematic diagram of the SAC305 dog-bone samples for creep testing with dimension labelled. The red line is drawn by diamond pen to thin the sample at back and create a local stress concentration, to localise microstructural damage (the volume of the back thin is negligible in engineering strain calculation). (b) Optical image of directionally solidified samples (initial microstructure) after β-Sn etching. Note that only directionally solidified samples were used for creep testing.*

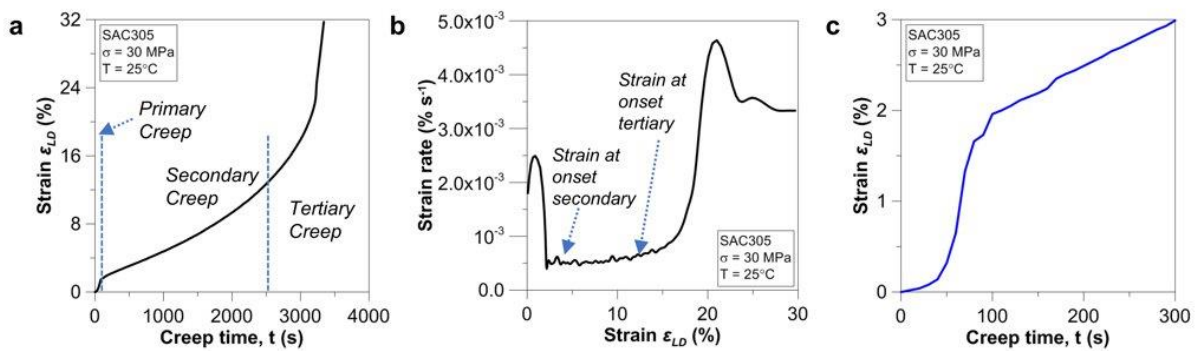

*Figure 2. Fixed load (30 MPa) mechanical creep data of SAC305 samples grown at v=20 μm/s. Creep curves for the bi-crystal sample (a, b) and for single crystal sample (c). (a) Creep strain vs. time showing the primary to tertiary stage of creep with three stages labelled. (b) Creep strain rate vs. creep strain curve shows a slope near 0 after the secondary stage creep is reached. The strain at the onset secondary and tertiary stages are labelled. (c) Creep curve of the single crystal sample until 3% of strain.*

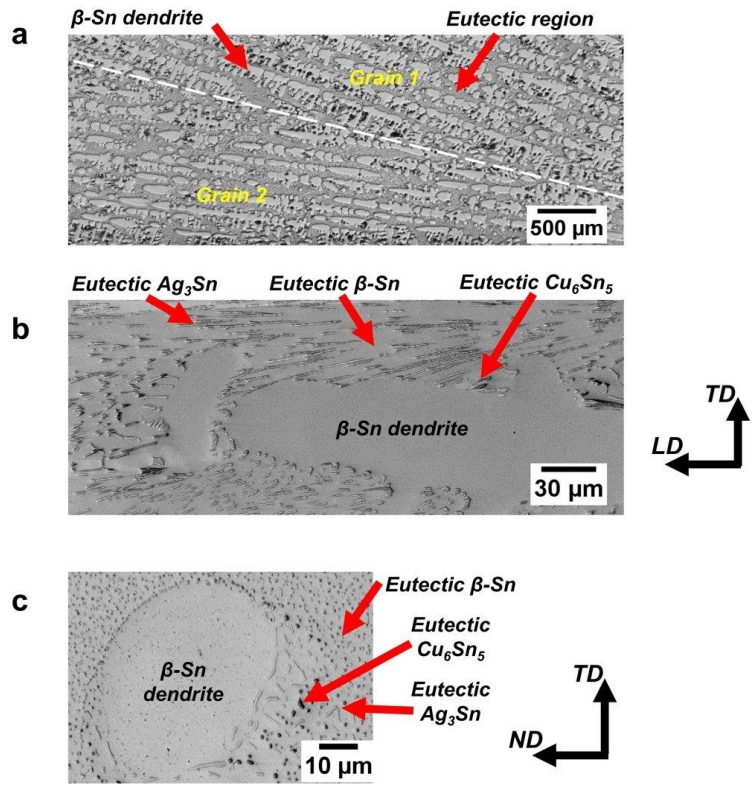

*Figure 3. (a) Optical image of the initial microstructure of the bi-crystal sample with β-Sn dendrite and eutectic regions labelled. The white dashed line indicates the grain boundary. (b, c) Backscattered electron images show the structure of β-Sn dendrite (b) and IMCs, e.g. $Ag_3Sn$ and $Cu_6Sn_5$ (c).*

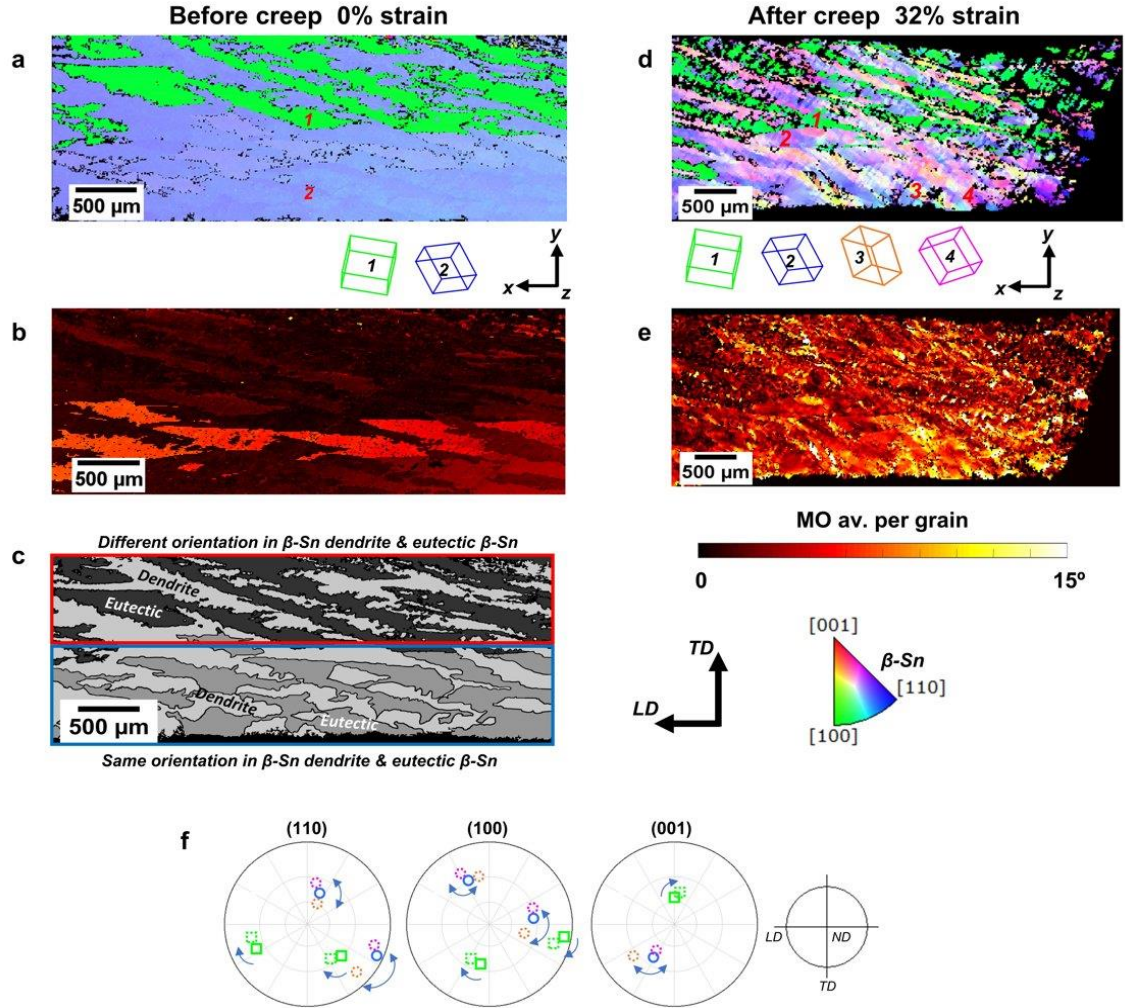

*Figure 4. Micrographs showing the evolution of the bi-crystal sample deformed under fixed load (30 MPa) and at room temperature (25 °C). (a – c) before creep at $\varepsilon_{LD} = 0\%$, (d, e) after creep at $\varepsilon_{LD} = 32\%$. (a, d) The EBSD crystal orientations are represented with inverse pole figure colouring with respect to the loading direction (IPF-LD) of the microstructure (only the β-Sn has been indexed here). (b, e) The misorientation shown as Grain Reference Orientation Deviation (GROD) maps. (c) The schematic of the initial microstructure of the sample showing the upper boxed region which has two different β-Sn orientations, with blue dendrite and green eutectic β-Sn, and the lower region where dendrites and eutectic have the same β-Sn orientation. The dendrite β-Sn is presented as light grey colour and eutectic β-Sn is shown as two darker grey colours. (f) The change in orientations of the sample is given in the pole figures (PF), where the initial green eutectic orientation is a green square and the deformed with green dashed square symbols. The initial orientation has blue circles and the corresponding deformed orientations are labelled with yellow (dendrite) and magenta (eutectic regions) dashed circles after fracture (d).*

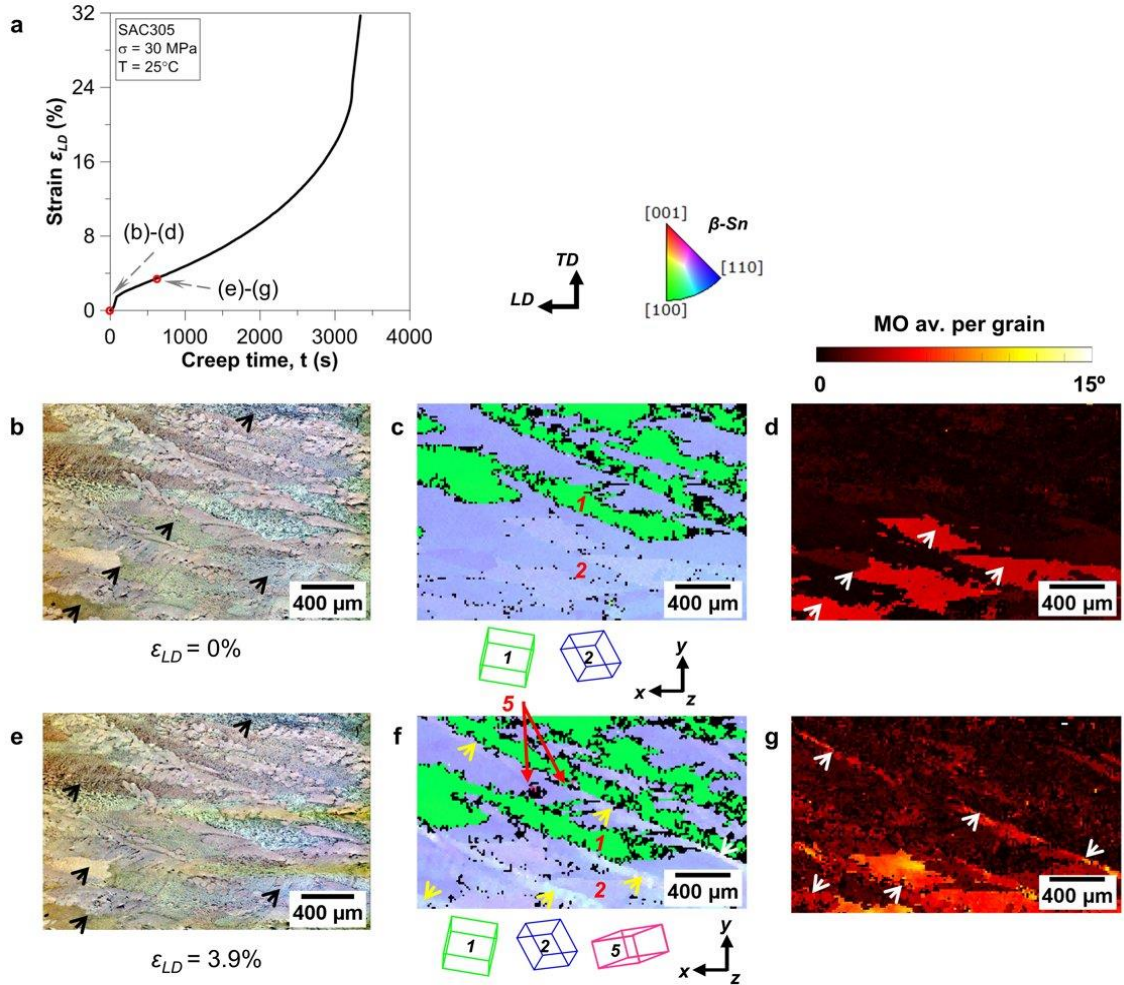

*Figure 5. EBSD crystal orientation maps showing microstructural evolution of the sample in early secondary stage creep (only the β-Sn has been indexed here). (a) Creep curve from primary to tertiary creep with mapping position labelled. (b-g) EBSD maps at initial condition ($\varepsilon_{LD}$ = 0%) (b-d), at early secondary stage ($\varepsilon_{LD}$ = 3.9%) (e-g). The initial high misorientation is presented within the grain 1 and subgrains of the grain 2 (d) (indicated with white arrows). The morphology of the dendrite and eutectic regions and contrast between the grains are shown in the FSD maps at the initial condition (b) and at the early secondary creep stage (e) where the subgrains are indicated with black arrows. With increasing strain, the orientation changes in the IPF-LD map (c, f) and misorientation increases in the GROD maps (d, g) in the strain localised regions, especially in the β-Sn dendrite near the grain boundaries and at the boundaries between dendrite and eutectic region (indicated with yellow arrows). The initial high misorientation is presented within grain 1 and subgrains of grain 2 (d) (indicated with white arrows).*

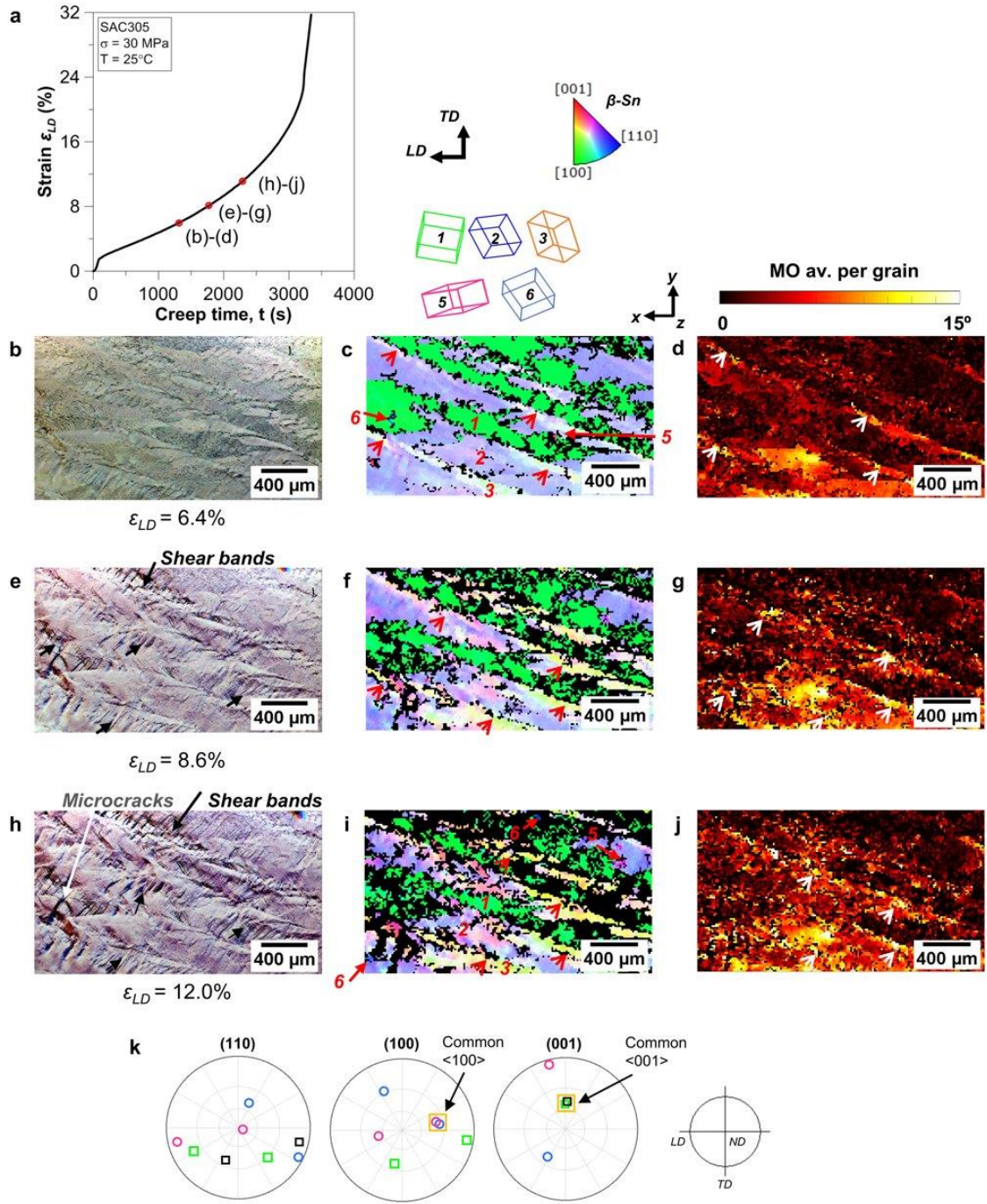

*Figure 6. EBSD crystal orientation maps showing microstructural evolution of a bi-crystal sample during secondary stage creep (only the β-Sn has been indexed here). (a) Creep curve from primary to tertiary creep with mapping strain labelled. (b-j) EBSD maps during secondary stage creep, at strain $\varepsilon_{LD} = 6.4\%$ (b-d), at strain $\varepsilon_{LD} = 8.6\%$ (e-g), and at strain $\varepsilon_{LD} = 12.0\%$ (h-j). With increasing strain, the surface roughness increases with formation of shear bands in the FSD maps (b, e h), the orientation changes with continuous subgrain formation in the IPF-LD map (c, f, i) and misorientation increases in the GROD maps (d, g, j) in the strain localised regions, especially at the boundaries between β-Sn dendrite and eutectic region and at the grain boundaries (indicated with red arrows).*

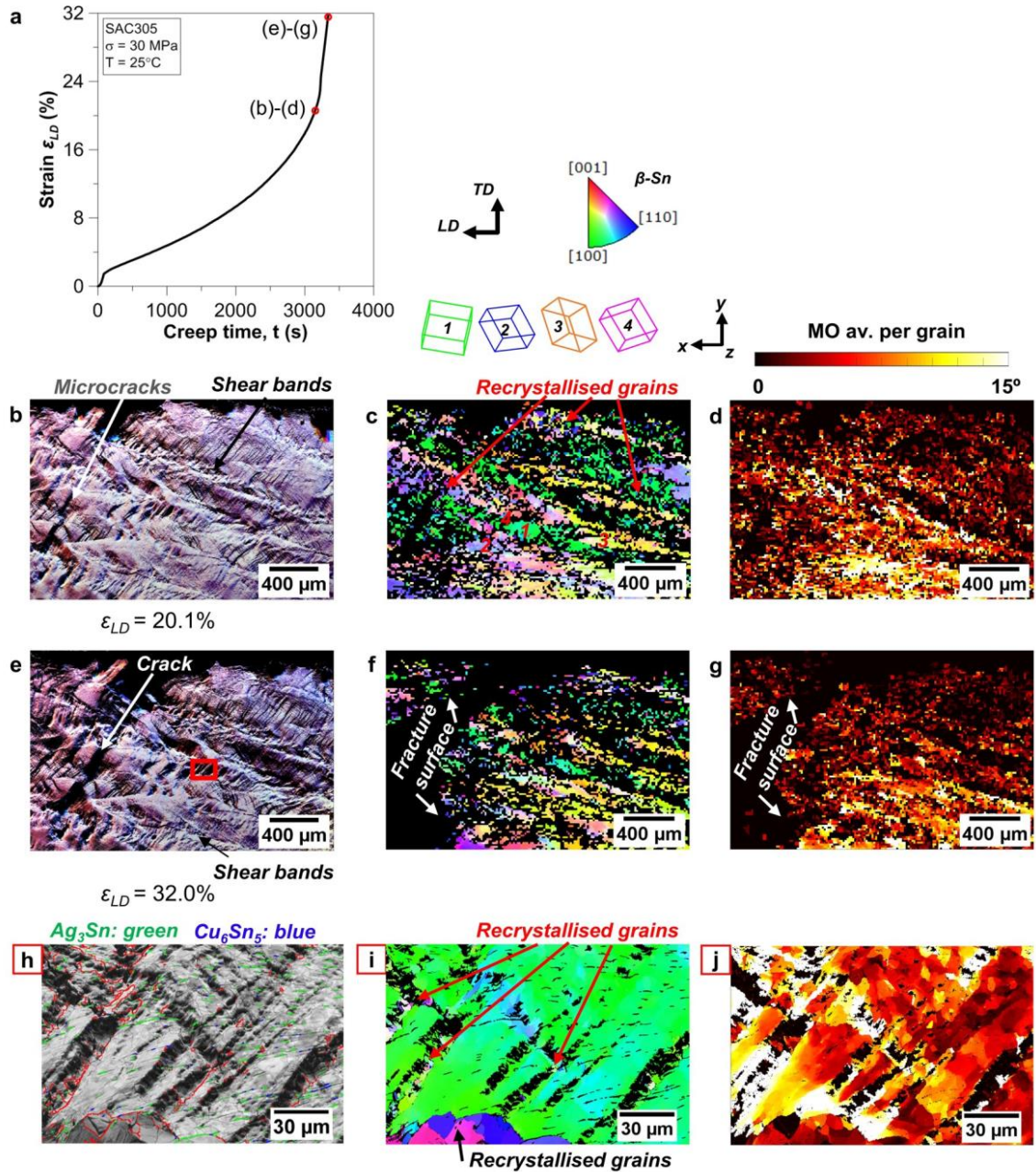

*Figure 7. EBSD crystal orientation maps showing microstructural evolution of a bi-crystal sample in tertiary stage creep (only the β-Sn has been indexed here). (a) Creep curve from primary to tertiary stage creep with mapping position labelled. (b-j) EBSD maps in tertiary stage creep, at strain $\varepsilon_{LD}$ = 20.1% (b-d), at strain $\varepsilon_{LD}$ =32.0% (e-g). With increasing strain, the surface roughness increases with crack propagation along the shear bands in the FSD maps (b, e), the orientation changes with continuous lattice rotation in the IPF-LD map (c, f,) and misorientation decreases in the GROD maps (d, g) due to release of stored energy (d, g). (h-j) Higher magnification maps of the red boxed region in (e), the high surface roughness is the eutectic region with $Ag_3Sn$ (green) and $Cu_6Sn_5$ (blue) identified and high-angle grain boundaries highlighted in red (h), subgrain formation and recrystallisation are shown in GROD map (j) and IPF-LD map (i) respectively.*

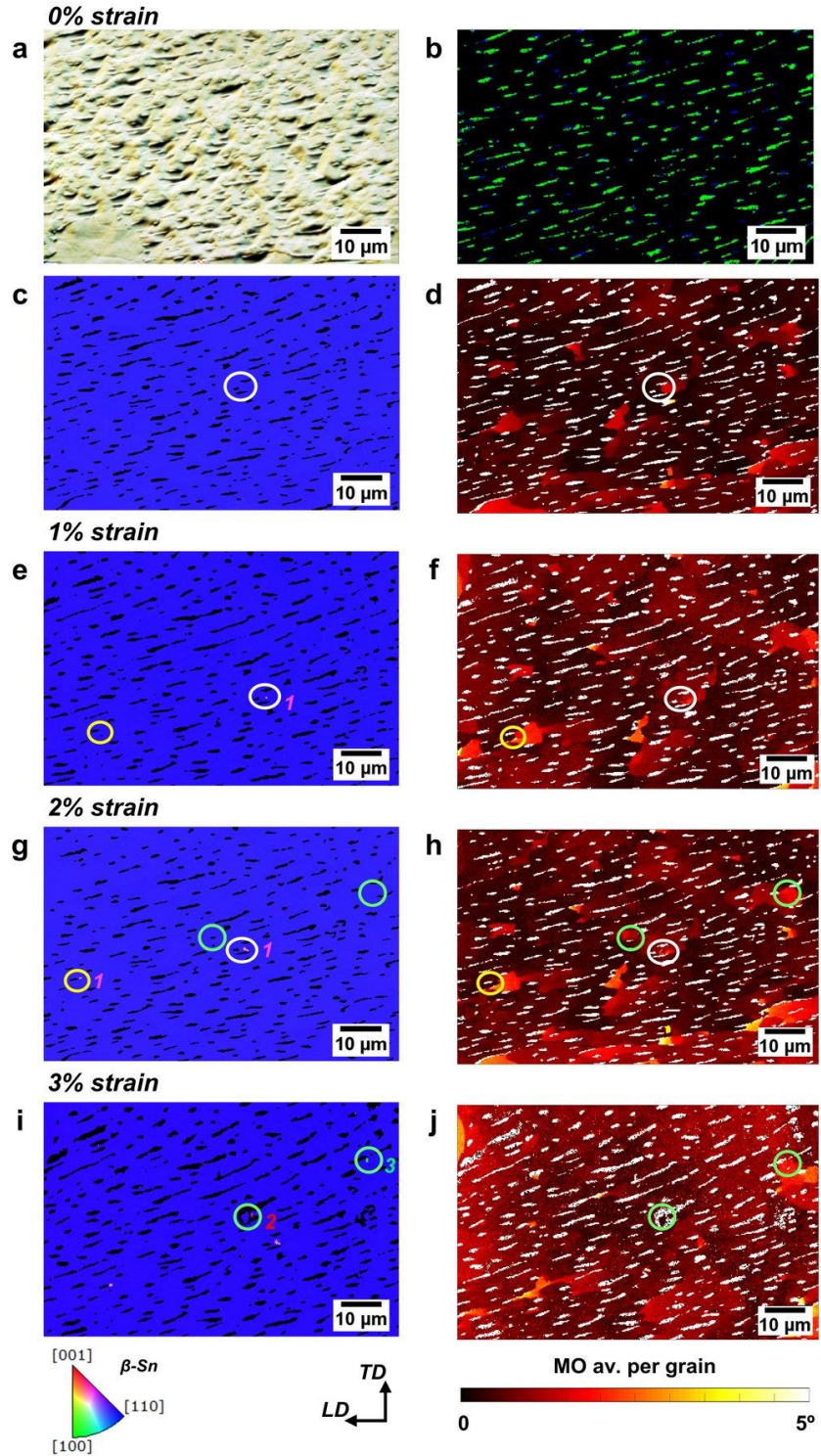

*Figure 8. (a-j) EBSD crystal orientation maps showing evolution of heterogeneous microstructure within a eutectic region of the single crystal sample. (a-d) at strain $\varepsilon_{LD} = 0\%$ (DS condition), (e, f) at strain $\varepsilon_{LD} = 1\%$, (g, h) at strain $\varepsilon_{LD} = 2\%$, (i, j) at strain $\varepsilon_{LD} = 3\%$. (a) The morphology of the eutectic regions is shown in the FSD map at initial condition. (b) The pattern quality map has $Ag_3Sn$ (green) and $Cu_6Sn_5$ (blue) identified. With increasing strain, the highlighted regions show formation of new recrystallised grains in the β-Sn at the β-Sn and IMCs interface in the IPF-LD maps (c, e, g, i) and gradual increase in misorientation across the mapped region with a slight decrease in local misorientation near the recrystallised grains in the GROD maps (d, f, h, j).*

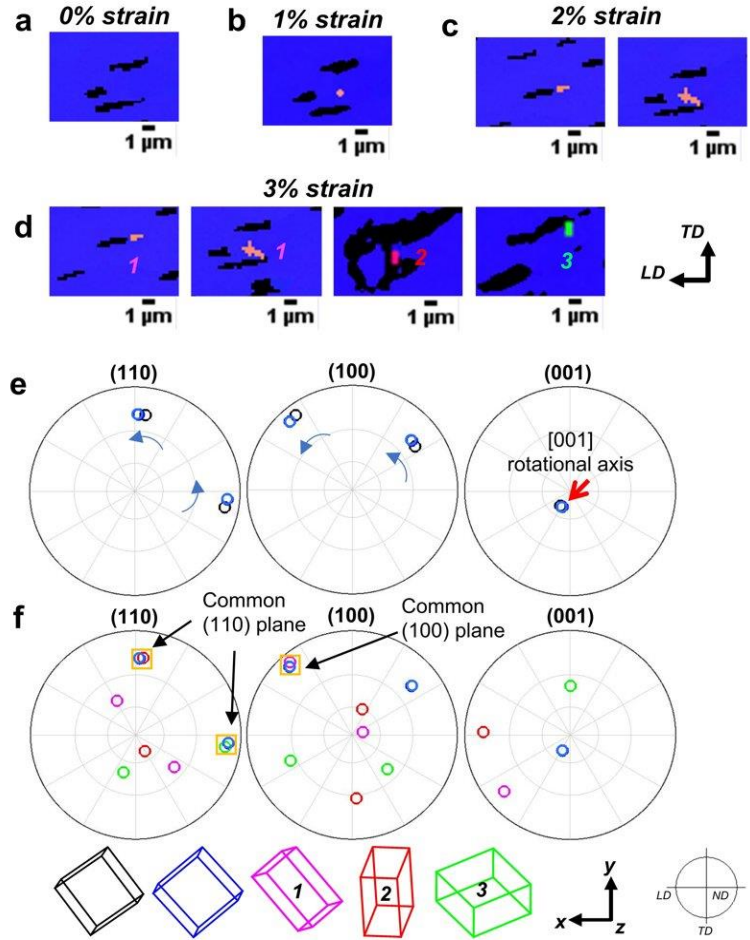

*Figure 9. (a-d) High magnification of IPF-LD maps around the circled regions in Figure 8.*

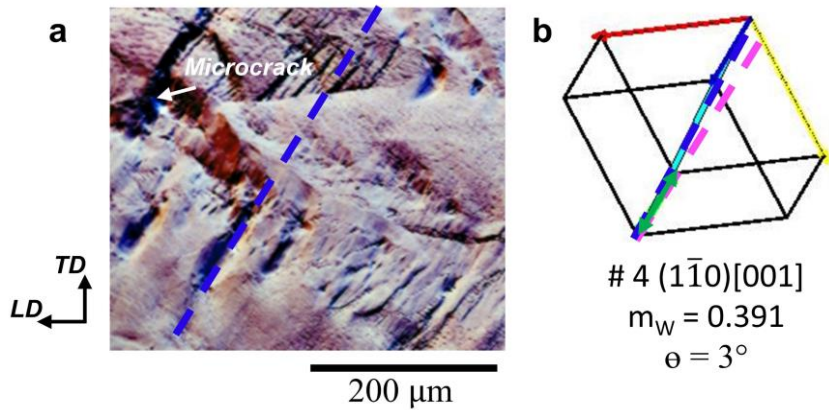

*Figure 10. (a) FSD image with the experimental slip trace labelled for the lower part of the examined region in Figure 4c. (b) The activated slip system shown on the unit cell with the highest weighted Schmid factor, $m_W$. The red, yellow and blue arrowed lines are the $a_1$, $a_2$, c crystal axes. The green arrowed lines are the slip directions. The cyan shaded plane represents the slip plane. The blue dashed line indicates the experimental slip trace and the calculated slip trace is indicated with the magenta dashed line, which differs by angle θ.*

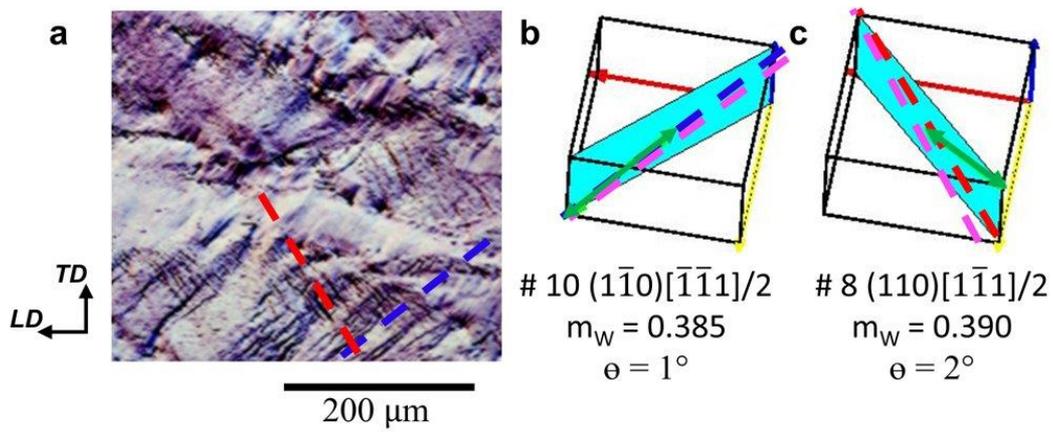

*Figure 11. (a) FSD image with 2 experimental slip traces labelled for the upper part of the examined region in Figure 4c. (b) The activated slip system shown on the unit cells with the highest weighted Schmid factor, $m_W$ for the 2 slip traces. The red, yellow and blue arrowed lines are the $a_1$, $a_2$, c crystal axes. The green arrowed lines are the slip directions. The cyan shaded planes represent the slip planes. The experimental slip traces are indicated with blue and red dashed lines and the calculated slip trace is indicated with magenta dashed lines, which differs by angle ө.*

**Supplementary Material**

| Family | Slip system | | CRSS ratio | CRSS ratio* | $m_w$ [110] | $m_w$ [100] |
|---|---|---|---|---|---|---|
| 1 | 1 | (100)[001] | 1 | 1.46 | 0.388 | 0.051 |
|   | 2 | (010)[001] | 1 | 1.46 | 0.163 | 0.012 |
| 2 | 3 | (110)[001] | 1 | 1 | 0.159 | 0.044 |
|   | 4 | (1$\bar{1}$0)[001] | 1 | 1 | 0.391 | 0.027 |
| 3 | 5 | (100)[010] | 1.05 | 2.54 | 0.263 | 0.218 |
|   | 6 | (010)[100] | 1.05 | 2.54 | 0.263 | 0.218 |
| 4 | 7 | (110)[1$\bar{1}$1]/2 | 1.1 | 1.77 | 0.280 | 0.361 |
|   | 8 | (110)[$\bar{1}$11]/2 | 1.1 | 1.77 | 0.176 | 0.390 |
|   | 9 | (1$\bar{1}$0)[111]/2 | 1.1 | 1.77 | 0.356 | 0.367 |
|   | 10 | (1$\bar{1}$0)[$\bar{1}\bar{1}$1]/2 | 1.1 | 1.77 | 0.100 | 0.385 |
| 5 | 11 | (110)[1$\bar{1}$0] | 1.2 | - | 0.224 | 0.369 |
|   | 12 | (1$\bar{1}$0)[110] | 1.2 | - | 0.224 | 0.369 |
| 6 | 13 | (010)[101] | 1.25 | 2.92 | 0.256 | 0.156 |
|   | 14 | (010)[10$\bar{1}$] | 1.25 | 2.92 | 0.131 | 0.165 |
|   | 15 | (100)[011] | 1.25 | 2.92 | 0.045 | 0.141 |
|   | 16 | (100)[01$\bar{1}$] | 1.25 | 2.92 | 0.342 | 0.180 |
| 7 | 17 | (001)[100] | 1.3 | - | 0.298 | 0.039 |
|   | 18 | (001)[010] | 1.3 | - | 0.126 | 0.009 |
| 8 | 19 | (001)[110] | 1.4 | - | 0.113 | 0.032 |
|   | 20 | (001)[1$\bar{1}$0] | 1.4 | - | 0.278 | 0.019 |
| 9 | 21 | (101)[10$\bar{1}$] | 1.5 | 1.23 | 0.259 | 0.245 |
|   | 22 | (10$\bar{1}$)[101] | 1.5 | 1.23 | 0.021 | 0.282 |
|   | 23 | (011)[01$\bar{1}$] | 1.5 | 1.23 | 0.091 | 0.010 |
|   | 24 | (01$\bar{1}$)[011] | 1.5 | 1.23 | 0.027 | 0.019 |
| 10 | 25 | (121)[$\bar{1}$01] | 1.5 | 1.15 | 0.111 | 0.272 |
|   | 26 | (1$\bar{2}$1)[$\bar{1}$01] | 1.5 | 1.15 | 0.262 | 0.082 |
|   | 27 | ($\bar{1}$21)[101] | 1.5 | 1.15 | 0.133 | 0.113 |
|   | 28 | ($\bar{1}\bar{2}$1)[101] | 1.5 | 1.15 | 0.163 | 0.293 |
|   | 29 | (211)[0$\bar{1}$1] | 1.5 | 1.15 | 0.263 | 0.111 |
|   | 30 | ($\bar{2}$11)[0$\bar{1}$1] | 1.5 | 1.15 | 0.132 | 0.096 |
|   | 31 | ($\bar{2}\bar{1}$1)[011] | 1.5 | 1.15 | 0.006 | 0.095 |
|   | 32 | (2$\bar{1}$1)[011] | 1.5 | 1.15 | 0.045 | 0.067 |

***Table S1.*** *The common Sn slip systems [21] with CRSS (the critical resolved shear stress) ratios given. The values of CRSS ratio are estimated from by Zamiri et al. [22] and the values of CRSS ratio\* are from Matin et al. [16]. The weighted Schmid factor ($m_W$) calculated from loading along the [110] ($m_w$ [110]) and [100] ($m_w$ [100]) orientations of the bi-crystal sample with CRSS ratio from ref [22]. Highlighting has been used to plot Figure 10 and Figure 11 in the manuscript.*

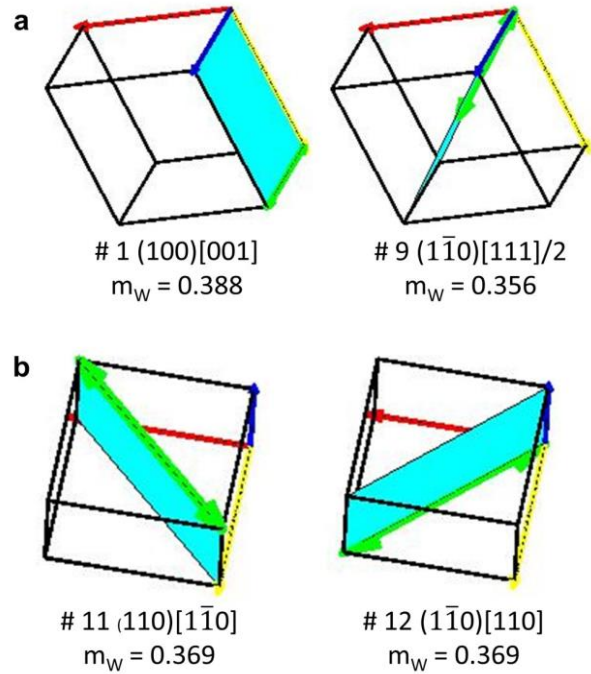

*Figure S1*. *The slip system shown on orientation unit cell with weighted Schmid factor, $m_W$ given for the bi-crystal sample (Figure 4a). (a) The near [110] crystal orientation (blue grain 2) with slip system #1 and 9 shown (underlined in Table S1), (b) the near [100] crystal orientation (green grain 1) with slip system #11 and 12 shown (underlined in Table S1). The red, blue yellow and blue arrowed lines are the $a_1$, $a_2$, c crystal axes. The green arrowed lines are the slip directions. The cyan shaded planes represent the slip planes.*

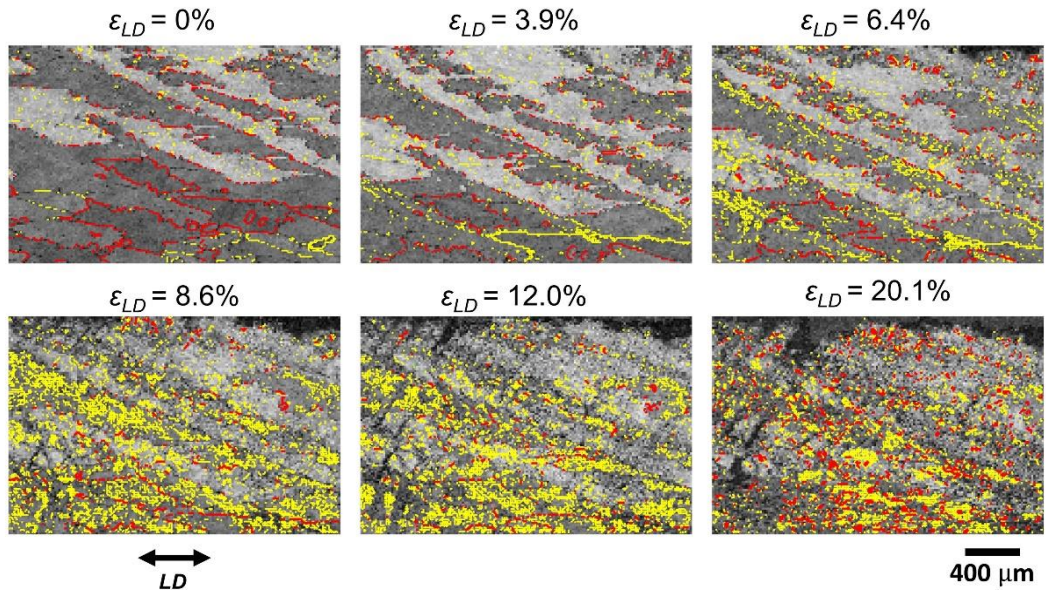

*Figure S2*. *EBSD pattern quality maps with high angle grain boundaries, HAGBs (indicated with red lines) and subgrain boundaries, LAGBs (indicated with yellow lines) overlaid at each strain level. The threshold of grain boundaries is identified as 5° and 2° for subgrain boundaries.*